# Design and optimization of a dispersive unit based on cascaded volume phase holographic gratings

Eduard R. Muslimov*[a,b], Gennady G. Valyavin[c], Sergei N. Fabrika[c], Nadezhda K. Pavlycheva[b]
[a]Aix Marseille Univ, CNRS, LAM, Laboratoire d'Astrophysique de Marseille, Marseille, France, 38 Frédéric Joliot-Curie, Marseille, France 13388; [b]Kazan National Research Technical University named after A.N. Tupolev-KAI 10 K. Marx, Kazan, Russian Federation 420111, [c]Special Astrophysical Observatory of the Russian AS, Nizhnij Arkhyz, Russian Federation 369167

## ABSTRACT

We describe a dispersive unit consisting of cascaded volume-phase holographic gratings for spectroscopic applications. Each of the gratings provides high diffractive efficiency in a relatively narrow wavelength range and transmits the rest of the radiation to the $0^{th}$ order of diffraction. The spectral lines formed by different gratings are centered in the longitudal direction and separated in the transverse direction due to tilt of the gratings around two axes. We consider a technique of design and optimization of such a scheme. It allows to define modulation of index of refraction and thickness of the holographic layer for each of the gratings as well as their fringes frequencies and inclination angles. At the first stage the gratings parameters are found approximately using analytical expressions of Kogelnik's coupled wave theory. Then each of the grating starting from the longwave sub-range is optimized separately by using of numerical optimization procedure and rigorous coupled wave analysis to achieve a high diffraction efficiency profile with a steep shortwave edge. In parallel such targets as ray aiming and linear dispersion maintenance are controlled by means of ray tracing. We demonstrate this technique on example of a small-sized spectrograph for astronomical applications. It works in the range of 500-650 nm and uses three gratings covering 50 nm each. It has spectral resolution of 6130-12548.Obtaining of the asymmetrical efficiency curve is shown with use of dichromated gelatin and a photopolymer. Change of the curve shape allows to increase filling coefficient for the target sub-range up to 2.3 times.

## INTRODUCTION

We consider an optical scheme of spectrograph based on a cascade of volume-phase holographic (VPH) gratings. The concept of such an instrument uses spectral selectivity and high diffraction efficiency which are inherent to the VPH's [1]. Each of the gratings in the cascaded dispersive unit provides high diffractive efficiency in a relatively narrow wavelength range and almost doesn't affect the beams propagation outside of this region. If a few of such gratings are mounted one after another with proper inclination angles in both of tangential and sagittal planes, it appears to be possible to form a spectral image consisting of several lines and characterized by a relatively high dispersion and high illuminance. Two versions of such a spectrograph were considered by the authors before [2,3].

One of possible problems with this type of design consists of partial overlapping and energy losses on the edges of the spectral sub-ranges. We must note here, that precise computation of the losses represents a separate modelling problem and should be considered separately. However, the losses can be minimized due to optimization of the diffraction efficiency (DE) curve for each grating. Specifically, in the present work we study possibilities to obtain a DE dependence with sharp edges and/or asymmetric profile, which could allow to decrease the losses and overlapping. The computation practice indicates, that DE curves with these features can be observed in cases when a conical diffraction is taken into account. Also, the properties of the holographic layer and the substrate have an influence on the actual DE curve. So we propose a design algorithm, in which optimization of the optical scheme geometry and the VPH gratings properties goes in parallel.

The paper is organized as follows. In section 2 we describe the algorithm used for design and optimization of the cascaded dispersive unit. In section 3 a certain example of the spectrograph optical scheme is shown and its' geometry and image quality are analyzed. Section 4 is devoted to the VPH gratings design analysis: we consider initial conditions and target functions used for the gratings parameters optimization, and demonstrate results obtained with different holographic materials. Section 5 contains discussion of the obtained results and proposals for their further application.

## DESIGN ALGORITHM

Previously optimization of the design in terms of image quality and resolution and that in terms of DE and throughput were performed separately. Groves frequency and angle of incidence in the tangential plane were varied to get the desired values of dispersion and provide longitudinal centering of the spectral images, while the gratings inclination angles in the sagittal plane were used for transverse separation of the image lines. On the other hand, such parameters as thickness of the holographic layer, refraction index modulation and slanting angle of the fringes were used to optimize the DE curve. But if additional targets like steepness of the DE curve are controlled, this separation is not possible anymore. Though the risk of overlapping exists only for grating, which are mounted in the cascade after the one after consideration. It is known from practice[4] that blueshift of a DE curve vertex is easier to obtain than its' redshift. So we should mount the longwave or 'red' grating on the first place in the cascade and then subsequently control steepness of the DE curve's 'blue' edge at each grating. With this assumption the optimization algorithm takes the form presented in Table 1.

Table 1. Algorithm of the VPH gratings cascade optimization.

| Step | Elements under consideration | Variable parameters in the optical scheme optimization | Variable parameters in the DE optimization |
|---|---|---|---|
| 1 | 'Red' grating, camera and collimator lenses | Camera and collimator focal length ($f'_{cam}=f'_{col}$) | Angle of incidence ($\theta_r$) Conical diffraction angle ($\delta_r$) Fringes frequency ($N_r$) Fringes slanting angle ($\beta_r$) Holographic layer thickness ($h_r$) Exposure ($E_r$) |
| 2 | 'Green' grating | Angle of incidence ($\theta_g$) Fringes frequency ($N_g$) | Conical diffraction angle ($\delta_g$) Fringes slanting angle ($\beta_g$) Holographic layer thickness ($h_g$) Exposure ($E_g$) |
| 3 | 'Blue' grating | Angle of incidence ($\theta_g$) Fringes frequency ($N_g$) Conical diffraction angle ($\delta_g$) | Fringes slanting angle ($\beta_g$) Holographic layer thickness ($h_g$) Exposure ($E_g$) |
| 4 | Camera and collimator lenses | Lenses design parameters | - |

As one can see from this table as the optimization moves along the optical path, the free parameters move from the DE targets to the optical scheme design targets. We must note here that the incident ray is defined by two spherical angles – angle with the normal ($\theta$) and angle between the incidence plane and the plane orthogonal to the fringes ($\delta$). We assume that the linear magnification between the collimator and camera is close to the unity and the entrance slit height is small enough to neglect it.

Further we consider in details optical design of a certain spectrograph scheme developed with use of this algorithm.

## OPTICAL DESIGN ANALYSIS

**Optical scheme**

The spectrograph works in the longwave visible domain 500-650 nm and uses 3 VPH gratings covering the sub-ranges of 500-550, 550-600 and 600-650 nm, respectively. The entrance beam has F/# of 4. It's collimated by a triplet-type lens with focal length equal to170 mm. The camera represents a double Gauss-type lens with the same focal length, which focuses the three spectral lines onto the detector plane. The length of each line is 25.6-26.5 mm and their separation is 0.9-1.1 mm. The optical scheme general view and the image format are shown on Fig.1.

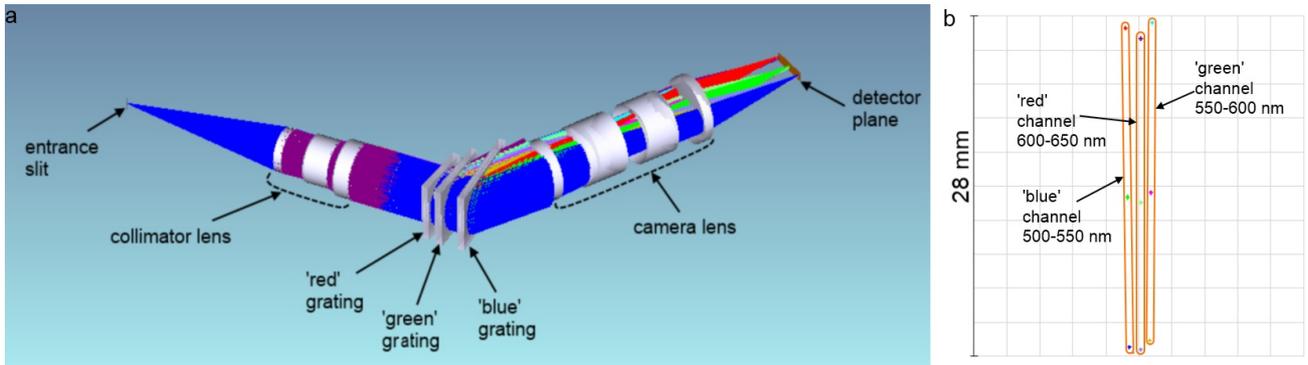

Figure 1.Optical design of the spectrograph with a cascade of VPH gratings: a – optical scheme, b – arrangement of the spectral images on the detector.

**Image quality**

For estimation of the image quality and resolution achieved in the scheme we use the spectrograph instrument function (Fig.2.) calculated for the entrance slit width of 25 μm.

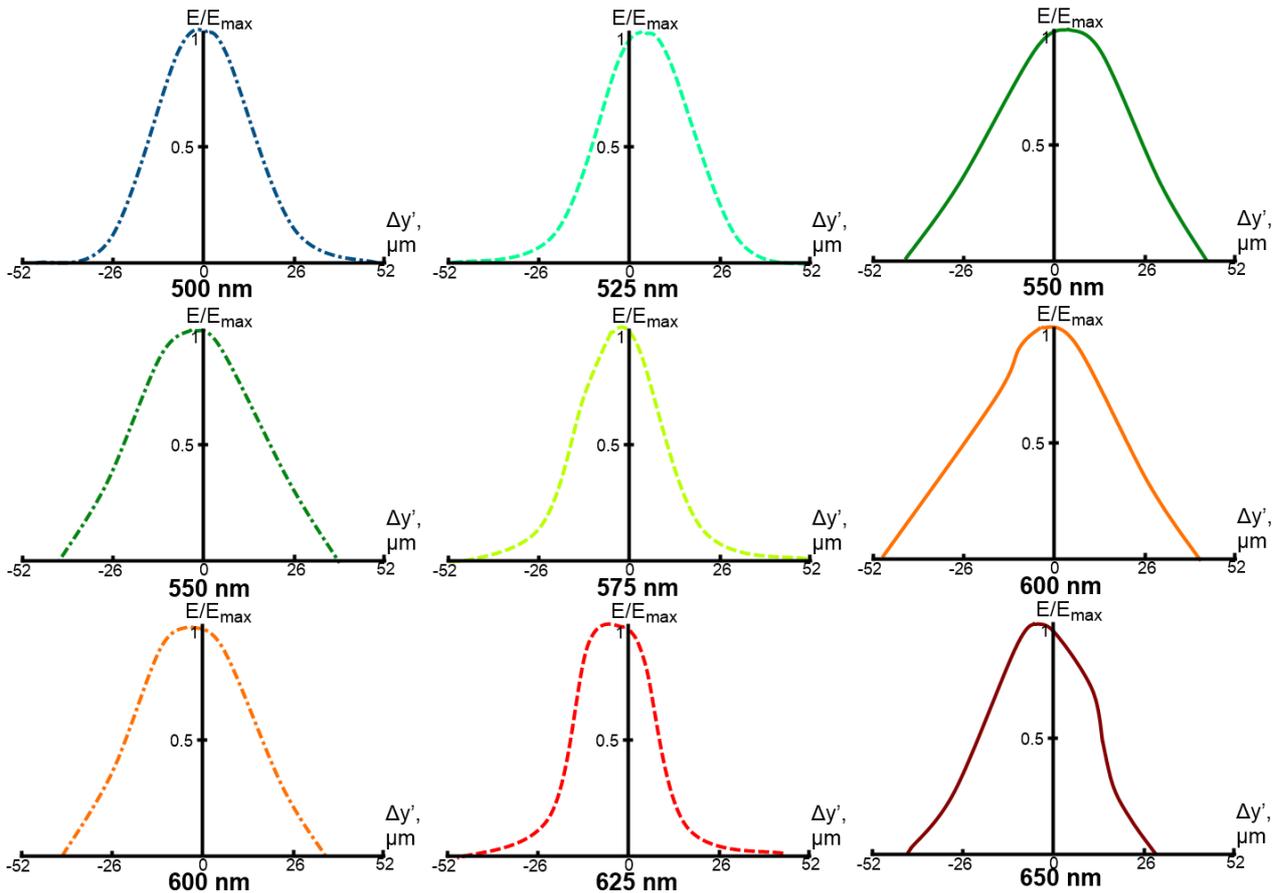

Figure 2. Instrument functions of the spectrograph calculated for the entrance slit width equal to 25 μm.

The instrument functions FWHM varies from 25.6 to 46.2μm across the spectrum. If we account for the reciprocal linear dispersion, which is 1.90-1.95 nm/mm for different wavelengths, we obtain spectral resolution equal to 6130-8651, 6776-11077 and 9491-12548 for the sub-ranges of 500-550, 550-600 and 600-650 nm, respectively.

## VPH GRATINGS DESIGN ANALYSIS

**Assumptions**

We assume that the gratings are VPH elements recorded on a sensitive layer embedded between substrate and cover plate made of BK7 glass. We account for the dispersive properties of the materials: Sellmeier formula[5] is used to model the glass and Schott formula[6] for the holographic layer.

When optimizing the holographic layer parameters, we do not use the refractive index modulation value directly. Instead we operate with value of exposure and apply a model of holographic material reaction. We consider mainly dichromated gelatin (DCG), the most widespread material for fabrication of VPH's. Its' properties are studied in details, so relying on the existing data[7] we model the DCG reaction in the following form:

$$\Delta n_1 = a \cdot \left(1 - e^{-\frac{E}{b}}\right)(c\lambda + d), \quad (1)$$

This equation indicates that the ascendant branch of the exposure reaction curve can be described by an exponent. In addition, the index modulation ($\Delta n_1$) depends linearly on the wavelength ($\lambda$). The coefficients values found for DCG are $a$=1.078, $b$=800, $c$= -0.00417 and $d$=0.01534.

Also for comparison we consider one of the modern holographic materials, namely the Bayfol™ photopolymer. Unfortunately, there is no data on change of its' dispersion with exposure, so we have to neglect it (i.e. $c$=0, $d$=1). The exposure reaction curve was studied[8,9] and if we consider only its' first part free of saturation, it's possible to use equation (1) with $a$=0.023; and $b$=0.087. Both of the models are presented graphically on Fig.3. Note that the exposure is given in different units, because of the difference in the initial data and the recording conditions.

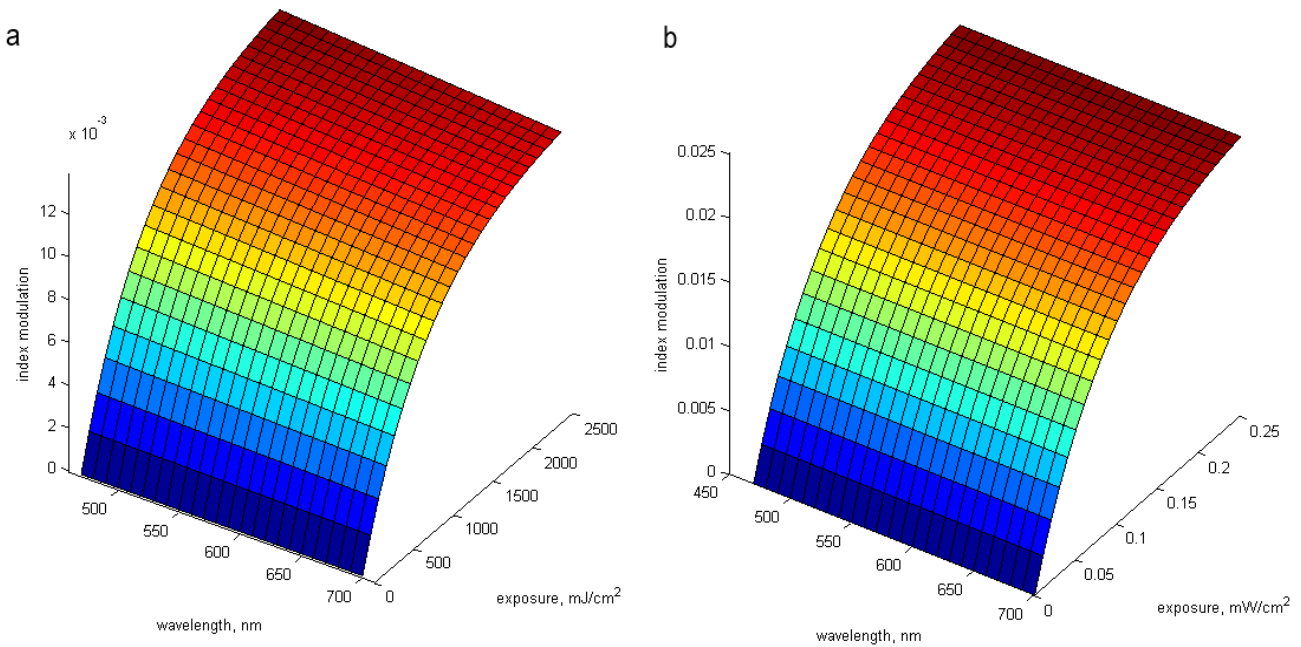

Figure 3. Dependencies of the refraction index modulation on exposure and wavelength: a – DCG, b – photopolymer.

In contrast with the approach used before[3], we put a special accent on steepness of the DE curve edge. So the target function used for optimization contains the curve inclination angle and DE value on a specified wavelength.

$$f_{tar}(x_i) = \left| \arctan\left( \frac{0.01}{\eta(\lambda_2) - \eta(\lambda_1)} \right) \right| + \frac{1}{\eta(\lambda_3)}, \qquad (2)$$

Here $x_i$ are the free parameters given in Table 1 for each case, $\eta$ is DE value on the specified wavelengths. For the simplicity reasons the $\lambda_{1-3}$ were taken 5 nm away from the sub-range borders (see Fig.4)

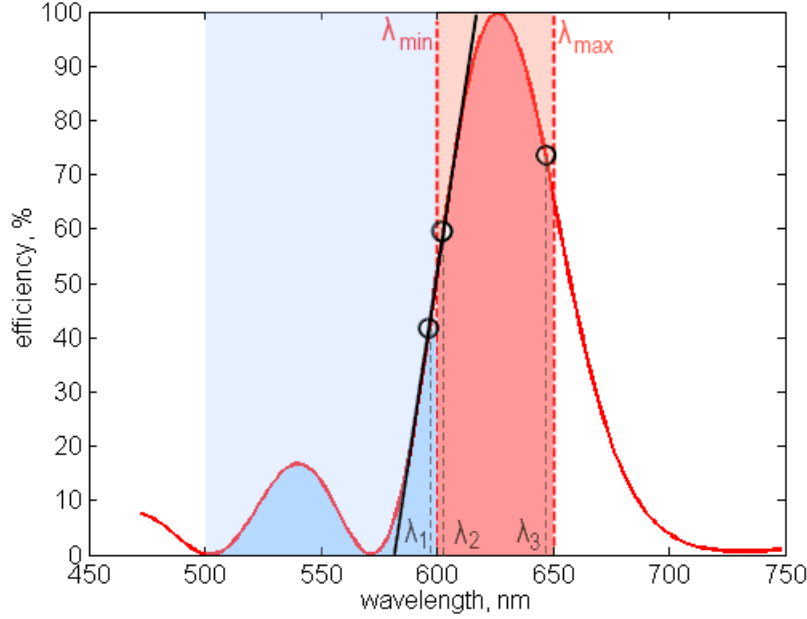

Figure 4. Diffraction efficiency curve of the 'red' VPH grating and parameters used for the target function description.

To assess the effect achieved due to optimization we integrate the DE curves over the working subrange and over the subranges of the gratings standing after it in the cascade. We obtain the following coefficients, which are approximately proportional to flux directed to the working sub-range and the losses, respectively. They correspond to the red and blue zones on Fig.4.

$$K_{in} = \int_{\lambda_{min}}^{\lambda_{max}} \eta(\lambda) d\lambda,$$
$$K_{ex} = \int_{500nm}^{\lambda_{min}} \eta(\lambda) d\lambda, \qquad (3)$$

To provide a simple and visual estimation of the achieved effect we also use the following coefficient, corresponding to filling of the rectangular areas (see the highlighted regions on Fig.4)

$$K_{fil} = \frac{1 \cdot (\lambda_{min} - 500) - K_{ex}}{1 \cdot (\lambda_{max} - \lambda_{min}) - K_{in}}, \qquad (4)$$

**Starting point**

To define the initial values of VPH parameters we use analytical equations of Kogelnik's coupled wave theory[10]. The DE curves obtained with these equations are shown on Fig. 5. The parameters values obtained after optimization are given in Table 3 in the end of this section together with those for other cases. It is easy to see that the DE curves have symmetrical profile, because the conical diffraction is set out of consideration by initial assumption of the theory. Subsequently, use of the target function described above leads to redshift of the curves maxima and decrease of DE in the working subrange.

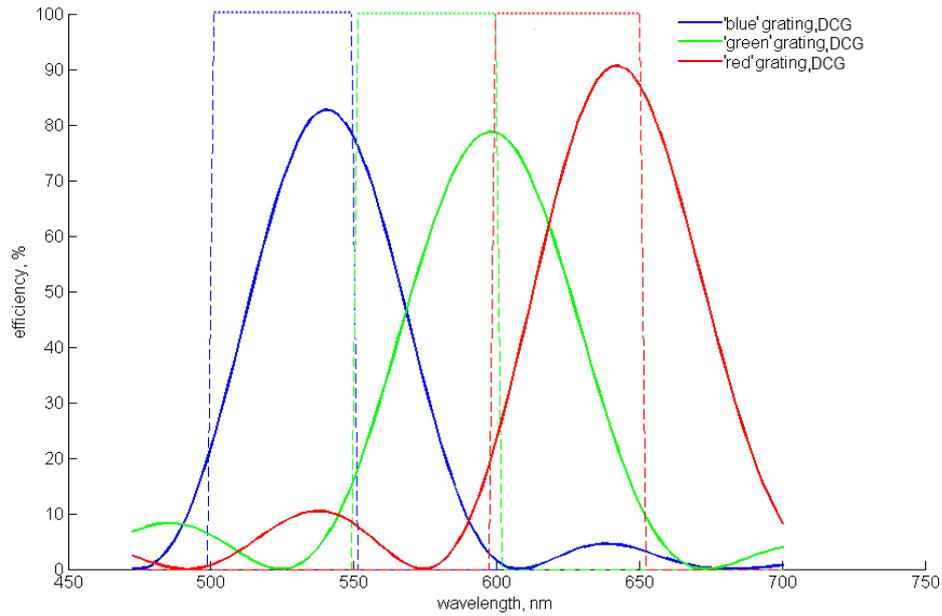

Figure 5. DE curves obtained with use of the Kogelnik's theory analytical equations.

**Optimized design for gratings recorded in DCG**

Using the starting point described above we perform optimization taking into account conical diffraction, change of the material dispersion with exposure. In this case rigorous coupled wave analysis (RCWA) method, implemented in *reticolo*[11,12] software is used. The resulting DE curves are shown on Fig.6, while the optimized values are summarized in Table 2.

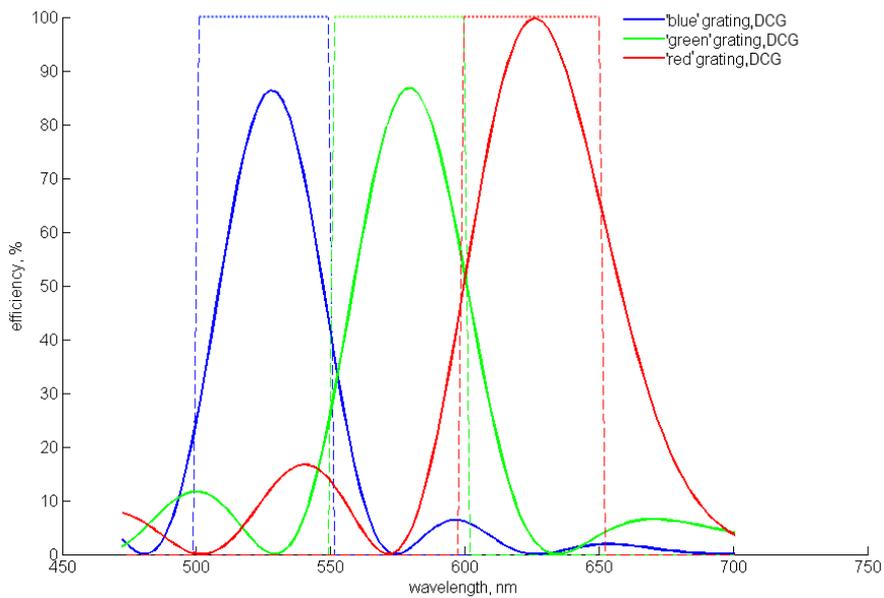

Figure 6. DE curves for gratings recorded in DCG obtained with use of RCWA method.

The plot demonstrates that using the conical diffraction one can obtain an asymmetric DE profile with blueshifted maximum and a steep shortwave edge. This effect is clearly visible for the 'red' grating while for the shorter wavelengths it gradually decreases. However, change of the curve shape also includes growth of its' secondary maxima, which can increase undesired overlapping of the sub-ranges and losses.

**Comparison with gratings recorded in photopolymer**

Finally, the DE optimization was performed with use of the photopolymer model. As well as in the previous cases, we provide the DE curves (Fig. 7) and optimized values of the gratings parameters (Table 2).

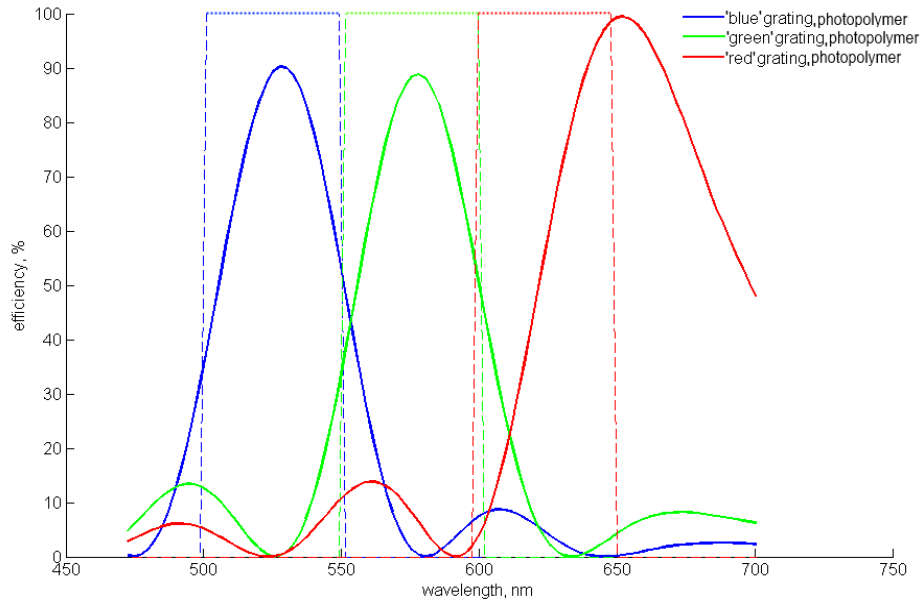

Figure 7. DE curves for grating s recorded in DCG obtained with use of RCWA method.

The results for 'blue' and 'green' gratings are similar to that obtained with DCG. The DE spectral dependence for the 'red' grating is characterized by a higher asymmetry, though the maximum position is shifted and therefore the total flux directed to the spectral image will decrease.

Table 2. Results of the gratings design optimization for different cases

| Parameter | Recording in DCG, Optimization with Kogelnik's equations | | | Recording in DCG, Optimization with RCWA | | | Recording in photopolymer, Optimization with RCWA | | |
|---|---|---|---|---|---|---|---|---|---|
| | blue | green | red | blue | green | red | blue | green | red |
| Angle of incidence $\theta$,° | 1.83 | 1.07 | 2.71 | 1.83 | 1.07 | 2.71 | 1.63 | 0.07 | 2.50 |
| Conical diffraction angle $\delta$,° | - | - | - | 17.57 | 20 | 0.02 | 17.52 | 19.99 | 0.29 |
| Fringes frequency $N$, $mm^{-1}$ | 1684 | 1478 | 1318 | 1684 | 1478 | 1318 | 1684 | 1478 | 1317 |
| Fringes slanting angle $\beta$,° | 15.5 | 14.33 | 12.66 | 15.50 | 15.67 | 13.18 | 15.7 | 16.6 | 13.99 |
| Holographic layer thickness $h$, $\mu m$ | 14 | 16 | 22 | 15.6 | 18.5 | 22.9 | 13.1 | 17.2 | 23.0 |
| Exposure $E$, $mJ/cm^2$ (*-$mW/cm^2$) | 2500 | 2500 | 2500 | 2500 | 2500 | 2500 | 0.143* | 0.097* | 0.070* |

Finally, to provide estimation of the obtained effect we summarize the coefficients defined by (3) and (4) in Table 3. The values indicate that use of an asymmetric DE curve allows to get better separation of the sub-ranges and higher irradiation in of the spectral image. The difference of the filling coefficients is as high as 2.3 times for the 'red' grating recorded in DCG and its' starting configuration with a symmetrical profile. Then for the 'green' and 'blue' gratings it becomes less notable (1.4 and 1.1 times, respectively). The results obtained with the photopolymer are more moderate. The gain obtained for the shorter wavelengths can be explained by larger achievable modulation values, while losses in the 'red' sub-range appears due to the maximum redshift. The latter feature can be compensated by revision of the target function or application of weight coefficients to it. In addition, we should mention here that such questions as concentration and developing conditions of the holographic materials are set out of consideration and used as they are given in the corresponding sources[7-9].

Table 3. Estimation of the gratings performance after optimization for different cases

| Parameter | Recording in DCG, Optimization with Kogelnik's equations | | | Recording in DCG, Optimization with RCWA | | | Recording in photopolymer, Optimization with RCWA | | |
|---|---|---|---|---|---|---|---|---|---|
| | blue | green | red | blue | green | red | blue | green | red |
| $K_{in}$ | 31.7 | 27.6 | 35.4 | 32.9 | 34.2 | 43.9 | 36.5 | 35.7 | 26.9 |
| $K_{ex}$ | 0 | 2.0 | 5.9 | 0 | 3.7 | 10.0 | 0 | 4.3 | 5.3 |
| $K_{fil}$ | 2.7 | 4.4 | 9.9 | 2.9 | 6.1 | 23.0 | 3.7 | 6.7 | 6.3 |

## CONCLUSION

We considered an algorithm of design and optimization of a spectrograph dispersive unit consisting of three cascaded volume-phase holographic gratings. The algorithm implies subsequent optimization of the gratings geometry and holographic layer parameters starting from the longwave sub-range. The gratings diffraction efficiency curves are controlled in parallel with their dispersive and imaging properties, while the free optimizations parameters are redistributed from the first group of target to the latter ones. Ray tracing is used when considering the dispersive and image properties. The DE optimization uses Kogelnik's theory for search of the starting point and RCWA method for accurate optimization.

Optical design of a small-sized spectrograph covering the range of 500-650 nm is considered as an example. It uses three VPH gratings providing uniform dispersion along the working range and allowing to reach spectral resolution from 6130 to 12548 depending on the wavelength. The gratings parameters are optimized to provide steep shortwave edges and maximum efficiency in the working sub-ranges. It is possible to obtain an asymmetric DE profile, thus increasing the spectral image illuminance and improving the channels separation. The best results were demonstrated for the 'red' grating and the filling coefficient in this case increases by factor of 2.3.

The achieved results can be of interest for further development of spectral instruments with VPH grating. Using the proposed approach one can achieve better performance in optical schemes with coupling of a few spectral or spectral and imaging channels.

The throughput of the best existing moderate and low-resolution spectrographs are 50 - 60 percent (for example, FOCAS/Subaru[13], OSIRIS/GTC[14], RSS/SALT[15]) and the efficiencies of even the best existing echelle-spectrometers HARPS[16] and PEPSI[17] still do not exceed 10 percent. Respectively, the low-moderate resolution spectrographs have spectral resolution of 1000 when only limiting optical range can be observed (at lesser resolution the whole range is covered, at higher resolution these can not cover all the visible range), the echelle-spectrographs may cover the whole range with spectral resolution from 10000 and notably higher. Respectively, in the first case one may observe faint objects, in the second, only much brighter.

This makes the use of the cascade-VPH spectrograph more effective in studies of faint, and/or low-contrast objects. The principal argument is that at the spectral resolution of 1000 or less, one may not distinguish between nebulae emission lines and all other objects, because the equivalent spectral resolution will be 300 km/s or worse. Nebula emission is overall in the universe and the nebulae have a dispersion velocity is less than 50 km/sec. The nebula emissions cover all the visible range, besides some principal lines like NaI $D_{12}$ lines must be resolved as well.

In such a case it is even impossible to distinguish between stellar winds of hot stars[18], accretion disks surrounding black holes or neutron stars[19, 20], cataclismic variables containing accretion disks with white dwars in close binaries systems, novae, peculiar novae, supernovae[21-23] and many other targets. It is because stellar winds or accretion disks have the dispersion about or higher than 100 - 200 km/sec, some narrow-line novae or supernovae have about the same values. That is crucially new to have spectral resolution higher than 6000 (50 km/sec), and therefore our solution is able to considerably contribute to modern areas of astrophysics.

Finally, the cascade-VPH effective spectrograph in combination with a 2-m class telescope can be used in studies of exoplanets. In particular, it will be very effective instrument in hunting for scatter light from hot-Jupiter exoplanets[24], and for spectral studies of transit events from giant planets[25]. These studies are presently among the hottest topics of modern astrophysics.

# ACKNOWLEDGEMENTS

This work was supported by the Russian Science Foundation (project No. 14-50-00043). Eduard Muslimov acknowledges the support of his personal post-doc contract by the European Research council through the H2020 - ERC-STG-2015 – 678777 ICARUS program. The authors also thank J.-P. Hugonin and Ph. Lalanne from Institut d'Optique for provision of the reticolo software used in this work.